# Direct observation of high temperature superconductivity in one-unit-cell FeSe films

Wenhao Zhang[1,3,#], Yi Sun[2,#], Jinsong Zhang[1], Fangsen Li[1,3], Minghua Guo[1], Yanfei Zhao[2], Huimin Zhang[3], Junping Peng[3], Ying Xing[2], Huichao Wang[2], Takeshi Fujita[4], Akihiko Hirata[4], Zhi Li[3], Hao Ding[1], Chenjia Tang[1,3], Meng Wang[3], Qingyan Wang[1], Ke He[1,3], Shuaihua Ji[1], Xi Chen[1], Junfeng Wang[5], Zhengcai Xia[5], Liang Li[5], Yayu Wang[1], Jian Wang[2,*], Lili Wang[1,3,*], Mingwei Chen[4], Qi-Kun Xue[1], and Xucun Ma[1,3]

[1]State Key Lab of Low-Dimensional Quantum Physics, Department of Physics, Tsinghua University, Beijing 100084, China

[2] International Center for Quantum Materials, School of Physics, Peking University, Beijing 100871, China

[3] Institute of Physics, Chinese Academy of Sciences, Beijing 100190, China

[4]WPI Advanced Institute for Materials Research, Tohoku University, Sendai 980-8577, Japan

[5]Wuhan National High Magnetic Field Center, Huazhong University of Science and Technology, Wuhan 430071, China

[#] Authors equally contributed to this work.

*Correspondence to: jianwangphysics@pku.edu.cn; liliwang@mail.tsinghua.edu.cn

**Heterostructure based interface engineering has been proved an effective method for finding new superconducting systems and raising superconductivity transition temperature ($T_C$)[1-7]. In previous work on one unit-cell (UC) thick FeSe films on SrTiO$_3$ (STO) substrate, a superconducting-like energy gap as large as 20 meV[8], was revealed by *in situ* scanning tunneling microscopy/spectroscopy (STM/STS). Angle resolved photoemission spectroscopy (ARPES) further revealed a nearly isotropic gap of above 15 meV, which closes at a temperature of 65 ± 5 K[9-11]. If this transition is indeed the superconducting transition, then the 1-UC FeSe represents the thinnest high $T_C$ superconductor discovered so far. However, up to date direct transport measurement of the 1-UC FeSe films has not been reported, mainly because growth of large scale 1-UC FeSe films is challenging and the 1-UC FeSe films are too thin to survive in atmosphere. In this work, we successfully prepared 1-UC FeSe films on insulating STO substrates with non-superconducting FeTe protection layers. By direct transport and magnetic measurements, we provide definitive evidence for high temperature superconductivity in the 1-UC FeSe films with an onset $T_C$ above 40 K and a extremely large critical current density $J_C \sim 1.7 \times 10^6$ A/cm$^2$ at 2 K. Our work may pave the way to enhancing and tailoring superconductivity by interface engineering.**

The FeSe films and FeTe protection layer are grown by molecular beam epitaxy (MBE) (see Methods). Figures 1a and 1b show typical STM topographic images of 1-UC FeSe film grown on insulating STO(001) substrate and that protected with a nominal 10-UC FeTe film, respectively. The 1-UC FeSe film exhibits atomically flat surface and regular steps originating from the STO substrate. The atomic-resolution STM topography image in Fig. 1a reveals that the in-plane lattice constant is 0.382 nm, slightly larger than the bulk value (0.377 nm) while still smaller than the in-plane lattice constant of STO(001) surface (0.391 nm), indicating small tensile strain in the 1-UC FeSe films. The FeTe protection layer grows on the FeSe film via a nearly layer-by-layer mode. At low coverage (< 4-UC), FeTe film covers the FeSe film uniformly like a carpet with regular steps originating from the STO substrate (Fig. S1). At the nominal thickness of 10-UC, the FeTe film is still quite flat except for some islands with single UC FeTe steps (0.63 nm) (Fig. 1b). The cross-sectional high-resolution scanning transmission electron microscope (STEM) image of the FeTe/FeSe/STO heterostructure (Fig. 1c) demonstrates atomically sharp interfaces of FeSe/STO and FeTe/FeSe with very few dislocations and defects. The line profile of the cross-sectional image (Fig. 1d) shows the out-of-plane lattice constant in FeTe layer is 0.63 nm, consistent with the STM results, and the 1-UC FeSe film is 0.55 nm thick, equal to the out-of-plane lattice constant in bulk FeSe[12,13]. Moreover, the mixed mapping of chemical distribution across the interfaces by Energy-dispersive X-ray spectroscopy (EDS) (Fig. 1e) indicates that there is basically no ion intermixing at the atomic scale. The FeTe protection layer is non-superconducting (Fig. S2 and S3) and a 30-nm-thick amorphous Si film is further deposited in order to avoid the possible superconductivity induced by oxygen incorporation into FeTe film[14]. Thus, the Si/FeTe/FeSe/STO heterostructure with atomically sharp FeTe/FeSe and FeSe/STO interfaces allows us to perform *ex situ* transport measurements of the 1-UC FeSe films.

The schematic for transport measurements is shown in the inset of Fig. 2a. Figure 2a is the resistance of the sample in *log* scale as a function of temperature, *R(T)*, at zero magnetic field (*B*) measured with the excitation current of 500 nA. The resistance starts to decrease at 54.5 K and drops completely to zero (defined as a resistance within the instrumental resolution of $\pm 0.04$ $\Omega$) at 23.5 K ($T_c^{zero}$). By extrapolating both the normal resistance and the superconducting transition curves, we obtain the onset $T_c^{onset}$ = 40.2 K, which is much higher than the $T_C$ ~ 8 K for bulk FeSe[12,13], and even higher than the $T_C$ ~ 36.7 K under hydrostatic high pressure of 8.9 GPa[15]. The diamagnetic response of these 1-UC FeSe films has been measured by a two-coil mutual inductance system[16]. As shown in Fig. 2b, the abrupt change of both the in-phase and out-of-phase signals at *T* ~ 21 K indicates the formation of diamagnetic screening, exactly corresponding to the zero resistance transition temperature. The superconducting transition is also confirmed by magnetic susceptibility measurements using a Magnetic Property Measurement System (Fig. S4). Compared with the previous transport result in 5-UC FeSe films (the resistance starts to decrease at

53 K but without exact zero resistance)[8], here the measurements directly show the high temperature superconductivity in 1-UC FeSe films and clearly reveal the zero resistance property.

Besides the high $T_C$, the 1-UC FeSe films can sustain superconductivity against high magnetic field up to 40 T. In order to determine the upper critical field $H_{C2}$, we carried out magneto transport at different temperatures by employing a pulsed high magnetic field up to 52 T. As shown in Fig. 3a, when the field is perpendicular to the film, the resistivity is significantly lower than the normal state value even at 52 T, unless for temperatures in the superconducting transition region. At 1.4 K, the resistance maintains zero up to 40 T and reaches a small fraction of the normal state value at 52 T, suggesting a much higher $H_{C2}$ value. When applying the magnetic field along the film plane, the upper critical field (Fig. 3b) is apparently higher than that in perpendicular field (Fig. 3a), which is consistent with the behavior of a typical two-dimensional (2D) superconductor.

More remarkably, the 1-UC FeSe thin films also exhibit a high critical current density $J_C$, a key factor of superconductors for practical applications. The $V(I)$ characteristics shown in Fig. 3c was measured at temperatures ranging from 2 to 50 K for zero field. The inset shows that $I_C$ is 13.3 mA at 0 T and 2 K. In our measurements, the STO substrate is insulating (Fig. S2) and neither FeTe protection layer nor STO substrate is induced superconducting by proximity effect (Fig. S3). Therefore, the superconductivity is confined to the 1-UC FeSe film with thickness of 0.55 nm. In Fig. 3d, we show the temperature dependence of $J_C$ calculated from $I_C$ under various magnetic fields. Below 12 K, the $J_C$ of the film is always larger than $1 \times 10^6$ A/cm$^2$ ($1.7 \times 10^6$ A/cm$^2$ at 2 K), which is nearly two orders of magnitude higher than that of bulk FeSe (~$10^4$ A/cm$^2$ at 1.8 K)[13]. Moreover, even under a 16 T magnetic field, the $J_C$ remains a large value of ~$10^5$ A/cm$^2$ below 8 K. The $J_C$ values are comparable to those of iron-chalcogenide-coated conductors[17] and textured $Sr_{1-x}K_xFe_2As_2$ tapes[18] reported very recently.

Further analysis on $V(I)$ characteristics reveals the signature of Berezinski-Kosterlitz-Thouless (BKT) transition, which is another evidence for 2D superconductivity[2,19,20]. As shown in Fig. 4a, measured $V(I)$ curves of the 1-UC FeSe films exhibit a $V \sim I^\alpha$ power-law dependence, and the slope corresponding to the exponent changes systematically as expected for BKT transition. A detailed evolution of α-exponent as a function of temperature is summarized in Fig. 4b. With decreasing temperature, the exponent α deviates from 1 and approaches 3 at 23.1 K, which is identified as $T_{BKT}$. This value is almost the same as $T_c^{zero}$ (23.5 K). Additionally, the observed $R(T)$ characteristics are consistent with a BKT transition, *i. e.* close to $T_{BKT}$, $R(T) = R_0 exp[-b(T/T_{BKT}-1)^{-1/2}]$, where $R_0$ and b are material parameters. As shown in Fig. 4c, $[dln(R)/dT]^{-2/3}$ vs. $T$ curve yields $T_{BKT}$ = 23.0 K, which is in agreement with the result of the $V$-$I^\alpha$ analysis.

In summary, we have unambiguously demonstrated the occurrence of high temperature superconductivity in 1-UC (0.55 nm) thick FeSe films on STO by direct transport and Meissner effect measurements. Owing to its simple structure, the current system offers an ideal platform for understanding the fundamental nature of unconventional superconductivity. The enhanced superconductivity with high $T_C$, $H_{C2}$ and $J_C$ makes the 1-UC FeSe thin films on high-dielectric STO substrate attractive for potential applications, such as superconducting interconnects, superconducting quantum interference device (SQUID), and field-effect transistor (FET) devices.

**Methods:**

The insulating single crystal STO(001) was pretreated by standard chemical etching with 10% HCl solution and thermal annealing in a tube furnace in order to obtain a specific $TiO_2$-terminated surface. It was then transferred into the ultra-high vacuum MBE chamber and annealed at 600 ℃ for 3 hours. Through the above treatments, the STO(001) surface becomes atomically flat with well-defined step-terrace structure. The 1-UC FeSe films were grown by co-evaporating Fe (99.995%) and Se (99.9999%) from Knudsen cells with a flux ratio of ~ 1:10 as the substrate was heated to 450 ℃ in the same MBE chamber. The growth rate is approximately 0.18 UC/min. After the growth, the FeSe films were gradually annealed up to 500 ℃ for several hours. The FeTe cover layers were grown on FeSe films by co-evaporating Fe (99.995%) and Te (99.9999%) with a flux ratio of ~ 1:4 on the FeSe films at 320 ℃. At last, a 30-nm-thick amorphous Si film was deposited on FeTe layer at 150 K for further protection. With or without the amorphous Si layer, the results on fresh samples are always same. With the amorphous Si layer, the superconductivity property of the films was found to degrade much slower and could survive for months. Our samples are rectangular strips around 7 mm long and 1.5 mm wide, limited by the size of sample holder in UHV-MBE-STM combined system (Omicron).

The atomic structure of the FeTe/FeSe/STO interfaces was characterized using a JEM-2100F TEM (JEOL, 200kV) equipped with a spherical aberration (Cs) corrector for the probe-forming lens. Elemental mappings were obtained using EDS. The mutual inductance is measured by a home-built two-coil system, and the sample is placed closely between the drive and pickup coils. During the measurement, an AC signal with amplitude of 5 μA and frequency of 10 kHz was applied to the driving coil, and the induced voltage in the pickup coil was measured by a lock-in amplifier. The out-of-phase signal (relative to the driving current) is the mutual inductance signal. It decreases rapidly below $T_C$ when the applied magnetic flux is partially repelled due to the Meissner transition. The in-phase signal is a second order mutual inductance signal produced by the eddy current in the film and detected by the pick-up coil.

**References:**

1. Strongin, M. *et al.* Enhanced superconductivity in layered metallic films. *Phys. Rev. Lett.* **21**, 1320-1323 (1968).
2. Reyren, N. *et al.* Superconducting interfaces between insulating oxides. *Science* **317**, 1196-1199 (2007).
3. Gozar, A. *et al.* High-temperature interface superconductivity between metallic and insulating copper oxides. *Nature* **455**, 782-785 (2008).
4. Kozuka, Y. *et al.* Two dimensional normal-state quantum oscillations in a superconducting heterostructure. *Nature* **462**, 487-490 (2009).

5. Pereiro, J., Petrovic, A., Panagopoulos, C. & Božović, I. Interface superconductivity: History, development and prospects. *Phys. Expr.* **1,** 208 (2011).

6. Zhang, T. et al. Superconductivity in one-atomic-layer metal films grown on Si (111). Nature Phys. 6, 104-108 (2010).

7. Uchihashi, T. Mishra, P., Aono, M. & Nakayama, T. Macroscopic superconducting current through a silicon surface reconstruction with indium adatoms: Si (111)-($\sqrt{7}\times\sqrt{3}$)-In. *Phys. Rev. Lett.* **107**, 207001 (2011).

8. Wang, Q.-Y. *et al.* Interface-induced high-temperature superconductivity in single unit-cell FeSe films on $SrTiO_3$. *Chin. Phys. Lett.* **29**, 037402 (2012).

9. Liu, D. *et al.* Electronic origin of high-temperature superconductivity in single-layer FeSe superconductor. *Nature Commun.* **3**, 931 (2012).

10. He, S. *et al.* Phase diagram and high temperature superconductivity at 65 K in tuning carrier concentration of single-layer FeSe films. *Nature Materials* **12**, 605-610 (2013).

11. Tan, S.-Y. *et al.* Interface-induced superconductivity and strain-dependent spin density wave in $FeSe/SrTiO_3$ thin films. *Nature Materials* **12**, 634-640 (2013).

12. Hsu, F. C. *et al.* Superconductivity in the PbO-type structure α-FeSe. *Proc. Natl. Acad. Sci. USA* **105**, 14262-14264 (2008).

13. Lei, H. C., Hu, R. W. & Petrovic, C. Critical fields, thermally activated transport, and critical current density of β-FeSe single crystals. *Phys. Rev. B* **84**, 014520 (2011).

14. Li, Q., Si, E. D. & Dimitrov, I. K. Films of iron chalcogenide superconductors. *Rep. Prog. Phys.* **74**, 124510 (2011).

15. Medvedev, S. *et al.* Electronic and magnetic phase diagram of β-$Fe_{1.01}$Se with superconductivity at 36.7 K under pressure. *Nature Mater.* **8**, 630-633 (2009).

16. Claassen, J. H. *et al.* A contactless method for measurement of the critical current density and critical temperature of superconducting films. *Rev. Sci. Instrum.* **62**, 996-1004 (1991).

17. Si, W. D. et al. High current superconductivity in $FeSe_{0.5}Te_{0.5}$-coated conductors at 30 tesla. *Nature*

*Commun.* **4**, 1347 (2012)

18. Gao, Z. et al. High critical current density and low anisotropy in textured $Sr_{1-x}K_xFe_2As_2$ tapes for high field applications. *Scientific Reports* **2**, 998 (2012).

19. Kosterlitz, J. M. & Thouless, D. J. Ordering, Metastability and phase-transitions in 2 dimensional systems. *J. Phys. C* **6**, 1181-1203 (1973).

20. Halperin, B. I. & Nelson, D. R. Resistive transition in superconducting films. *J. Low Temp. Phys.* **36**, 599 (1979).

**Acknowledgments**

The authors would like to thank Fa Wang, Qian Niu and Yuan Li for fruitful discussions. We are grateful to Ministry of Science and Technology and National Science Foundation of China and Chinese Academy of Sciences for financial supports.

**Additional information**

The authors declare no competing financial interests.

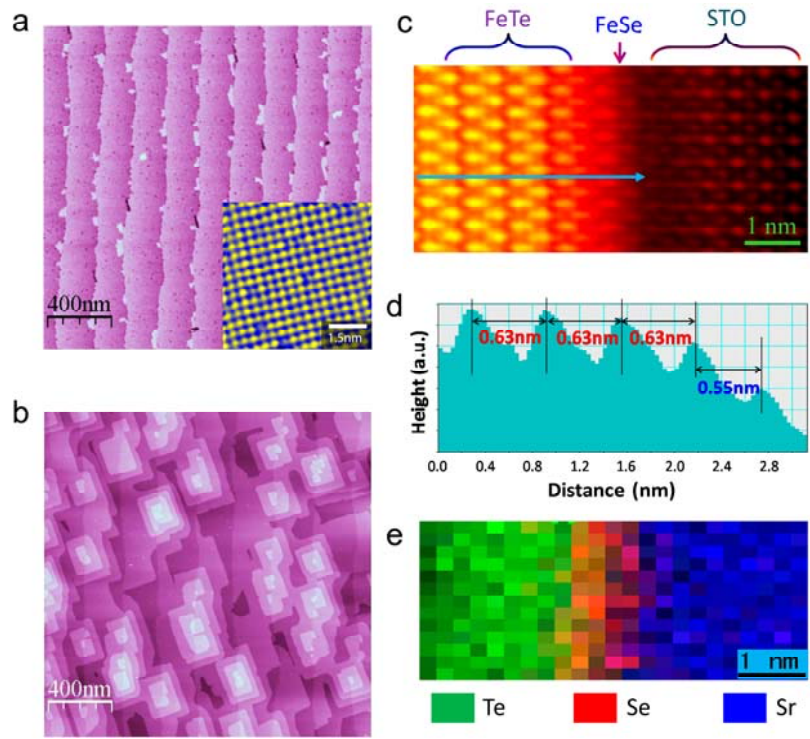

**Figure 1| STM and STEM observations of 1-UC thick FeSe films grown on SrTiO$_3$ (001). a**, A typical STM image of 1-UC FeSe films (image size 2000 nm × 2000 nm, sample bias V$_S$ = +2.0 V, tunneling current I$_t$ = 80 pA) with the atomic-resolution image inset (6 nm × 6 nm). The FeSe (001) surface is Se-terminated, and the in-plane lattice is ~ 0.382 nm. **b**, STM image of a 10-UC FeTe/1-UC FeSe/STO heterostructure (image size 2000 nm × 2000 nm, sample bias V$_S$ = +1.0 V, tunneling current I$_t$ = 60 pA). The steps are single unit-cell steps of FeTe (a Te-Fe-Te triple layer, 0.63 nm). **c**, The high-angle annular dark field STEM image of a 10-UC FeTe/1-UC FeSe/STO heterostructure, showing two very sharp interfaces. **d**, The line profile corresponding to the blue line in c, confirming the distinct FeTe, FeSe films and SrTiO$_3$ substrate. **e**, The mixed mapping of chemical composition across the interfaces by EDS, showing the spatial distribution of FeTe, FeSe and STO.

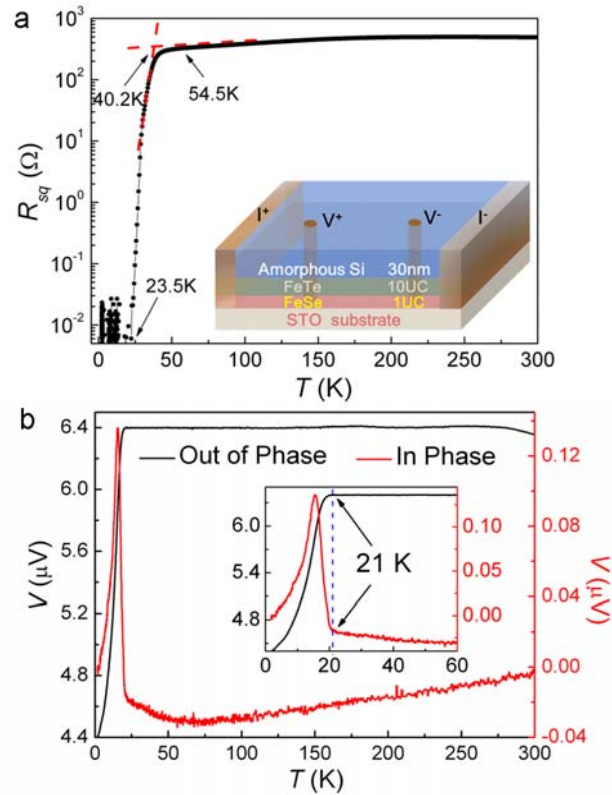

**Figure 2| Transport and diamagnetic measurements of 1-UC FeSe films grown on SrTiO$_3$ (001). a**, The temperature dependence of resistance under zero field, showing $T_c^{onset}$ = 40.2 K and $T_c^{zero}$ = 23.5 K. Inset: A schematic structure for the transport measurements in the heterostructure of 30 nm amorphous Si/10-UC FeTe/1-UC FeSe/STO. **b**, The diamagnetic response measured by a homebuilt two-coil mutual inductance system. Inset: The data near the superconducting transition in an amplified view showing the formation of diamagnetic screening at 21 K.

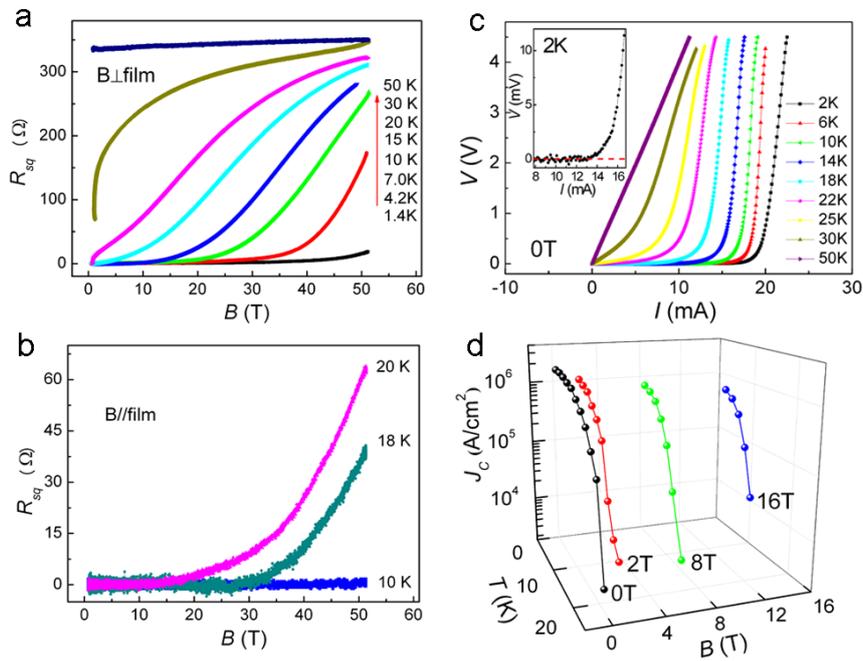

**Figure 3| Magnetoresistance, *V(I)* characteristics and critical current density ($J_C$) of 1-UC FeSe films grown on SrTiO$_3$ (001). a**, Magnetoresistance measured in perpendicular magnetic field by utilizing pulsed magnetic field up to 52 T. **b**, Magnetoresistance measured in parallel magnetic field. **c**, *V(I)* curves measured at temperatures ranging from 2 to 50 K for B = 0 T. **d**, $J_C$ calculated from $I_C$ at various temperatures and perpendicular magnetic fields.

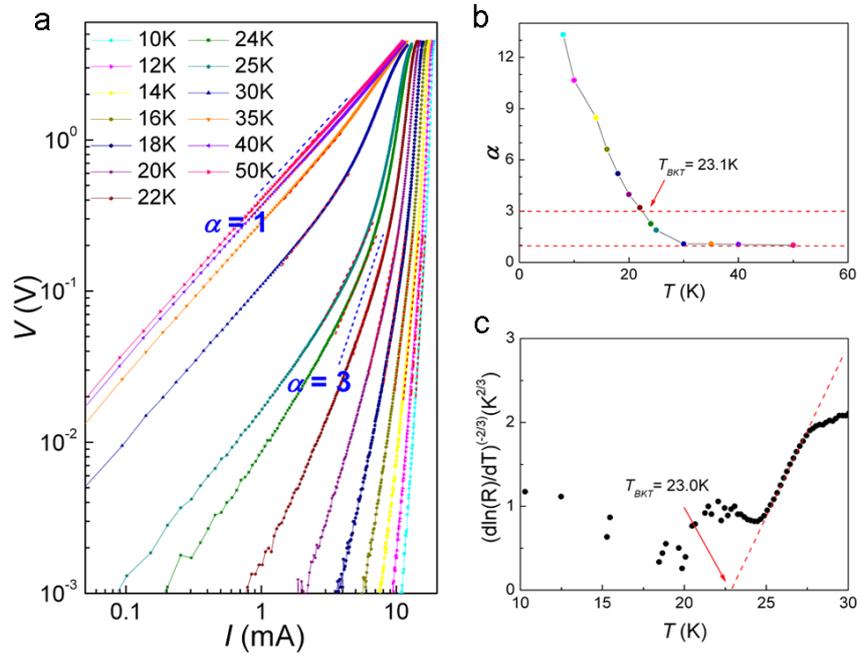

**Figure 4| Berezinski-Kosterlitz-Thouless transition of 1-UC FeSe films grown on SrTiO$_3$ (001). a**, *V(I)* characteristics at various temperatures plotted on a logarithmic scale. The two dashed blue lines correspond to *V ~ I* and *V ~ I$^3$* dependences, respectively. **b**, The variation of exponent *α* as a function of temperature, extracting from the power-law fittings in a panel, showing that $T_{BKT}$ = 23.1K. **c**, The *R(T)* curve plotted with the *[dln(R)/dT]$^{2/3}$* scale. The dashed line shows the fitting to the Halprin-Nelson formula $R(T) = R_0 exp[-b(T/T_{BKT}-1)^{-1/2}]$ with $T_{BKT}$ = 23.0 K.

Supplementary Information for

# Direct observation of high temperature superconductivity in one-unit-cell FeSe films

Wenhao Zhang[1,3,#], Yi Sun[2,#], Jinsong Zhang[1], Fangsen Li[1,3], Minghua Guo[1], Yanfei Zhao[2], Huimin Zhang[3], Junping Peng[3], Ying Xing[2], Huichao Wang[2], Takeshi Fujita[4], Akihiko Hirata[4], Zhi Li[3], Hao Ding[1], Chenjia Tang[1,3], Meng Wang[3], Qingyan Wang[1], Ke He[1,3], Shuaihua Ji[1], Xi Chen[1], Junfeng Wang[5], Zhengcai Xia[5], Liang Li[5], Yayu Wang[1], Jian Wang[2,*], Lili Wang[1,3,*], Mingwei Chen[4], Qi-Kun Xue[1], and Xucun Ma[1,3]

[1]*State Key Lab of Low-Dimensional Quantum Physics, Department of Physics, Tsinghua University, Beijing 100084, China*

[2]*International Center for Quantum Materials, School of Physics, Peking University, Beijing 100871, China*

[3]*Institute of Physics, Chinese Academy of Sciences, Beijing 100190, China*

[4]*WPI Advanced Institute for Materials Research, Tohoku University, Sendai 980-8577, Japan*

[5]*Wuhan National High Magnetic Field Center, Huazhong University of Science and Technology, Wuhan 430071, China*

For transport measurement, tens of samples have been measured and they showed similar results. We chose indium as electrodes and pressed it into the film firmly so that indium electrodes can penetrate through the Si and FeTe protection layers. We placed two current electrodes (I+ and I-) on both ends and across the entire width of the film strip with a size of *0.3 mm×1.5 mm*, so that the current can homogeneously go through the FeSe film in longitudinal direction. The diameter of the indium voltage electrodes is about 0.3 mm and the distance between them is about 1.8 mm. The *R(T)* and *R(H)* curves were measured on a commercial Quantum Design PPMS with the magnetic field up to 16 T (PPMS-16 system) with an excitation current of 500 nA. The *V(I)* curves were measured on the PPMS-16 system with an electrical transport option (ETO-P605) by varying current in both directions (low to high and vice versa), which yield almost identical results. Therefore, we only used the data collected when the current was scanned upward. The values of voltage are in the range of several volts during scanning current within tens of mA. The samples have never been broken down under such condition. The high-field magnetoresistance was measured by using an 80 ms non-destructive pulsed magnet. Open loops in the measuring circuit were minimized by twisting wires. To further eliminate spurious signals in pulsed fields, the magnetoresistance was measured by comparison of two pulse shots with different bias currents.

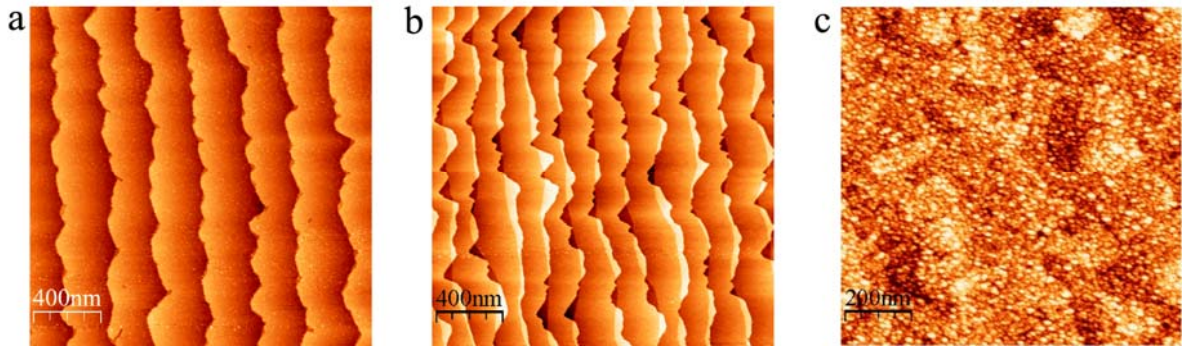

**Figure S1| STM images of Si/FeTe/FeSe/STO heterostructured film. a**, A typical STM topography of insulating SrTiO$_3$ (001) substrate (image size 2000 nm × 2000 nm, sample bias V$_S$ = +3.0 V, tunneling current I$_t$ = 60 pA). All the step heights are 0.4 nm, indicating single (TiO$_2$) termination. **b**, STM image of a 2-UC FeTe/1-UC FeSe/STO heterostructure (image size 2000 nm × 2000 nm, sample bias V$_S$ = +1.0 V, tunneling current I$_t$ = 60 pA). The 2-UC FeTe films covered the 1-UC FeSe film uniformly. The steps are 0.4 nm, originating from the STO substrate. **c**, STM image of a 30-nm Si/10-UC FeTe/1-UC FeSe/STO heterostructure (image size 1000 nm × 1000 nm, sample bias V$_S$ = +3.0 V, tunneling current I$_t$ = 60 pA). The Si film covers the FeTe film in the form of amorphous clusters.

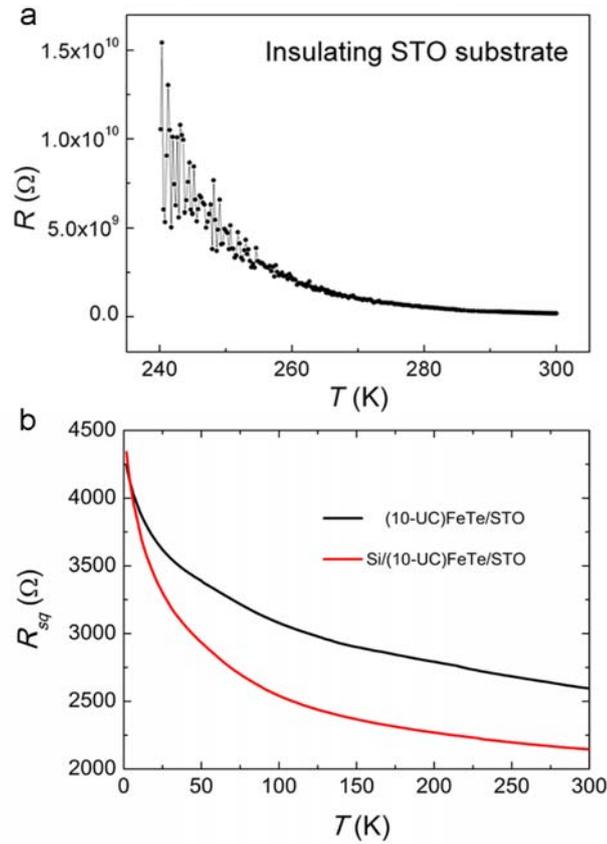

**Figure S2| Insulating-like property of STO substrate and FeTe films. a**, Temperature-dependent resistance of the STO substrate after being annealed at 600 ℃ for 3 hours in UHV. The resistance at 300 K is $1.8\times10^8$ Ω, and increases to $1.2\times10^{12}$ Ω when the temperature decreases to 240 K. When the temperature further decreases below 240 K, the value of resistance is too large to be detected by our instrument. **b**, Temperature-dependent resistance of a 10-UC thick FeTe film grown on STO substrate with and without Si capping layers, both showing the non-superconducting behavior down to 1.8 K.

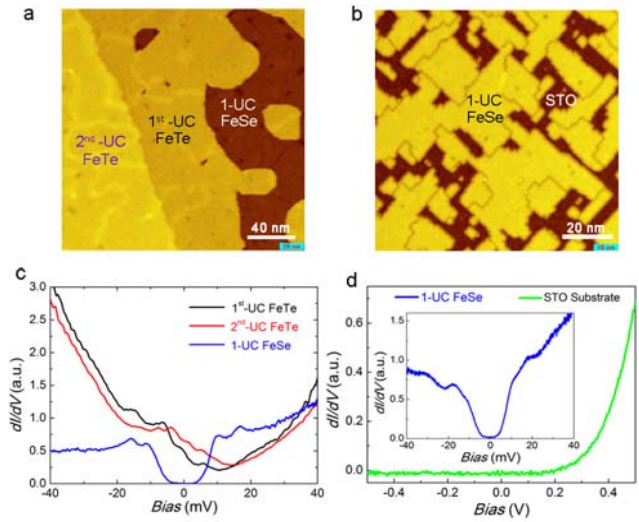

**Figure S3| Non-superconducting property of FeTe films and STO surface. a**, The STM image of 1-UC FeSe film partially covered by FeTe protection layer (image size 200 nm × 200 nm, sample bias $V_S$ = +1.5 V, tunneling current $I_t$ = 30 pA). The surface consists of 1-UC FeSe, 1st-UC FeTe and 2nd-UC FeTe. **b**, The STM image of sub-UC FeSe films on STO substrate. The surface consists of 1-UC FeSe films and exposed STO surface (100 nm × 100 nm, $V_S$ = 1.5 V, $I_t$ = 30 pA). **c**, The typical STS taken on 1st-UC FeTe, 2nd-UC FeTe and 1-UC FeSe films in a, showing non-superconducting behavior of FeTe protection layer. **d**, The typical STS taken on 1-UC FeSe films and exposed STO surface in b, showing non-superconducting behavior of STO substrate. Here, the substrate is Nb-doped STO(001).

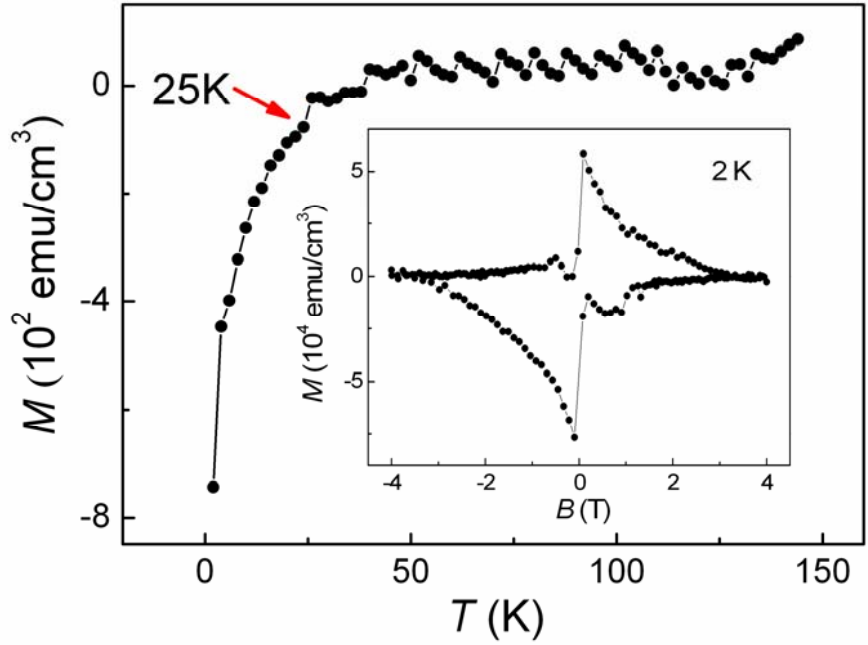

**Figure S4| Magnetic susceptibility of 1-UC FeSe films.** Temperature dependence of magnetic susceptibility for H = 1000 Oe parallel to the film shows a sharp drop around 25 K, which is consistent with the $T_c^{zero}$ observed in *R(T)* results (Fig. 2). The inset shows a typical magnetic hysteresis behavior measured at 2 K, which confirms the superconducting characteristics of the 1-UC FeSe film. It is measured by a Magnetic Property Measurement System (MPMS-SQUID-VSM) from Quantum Design Company.